\documentclass[prb,amsmath,amssymb,floatfix,twocolumn,citeautoscript]{revtex4}
\usepackage{bm}
\usepackage{epsfig}

\begin{document}

\title{Negative-$U$ Superconductivity on the Surface of Topological Insulators}

\author{Jian-Huang She$^{1, 2}$, Alexander V. Balatsky$^{3, 4}$}

\affiliation{$^1$Department of Physics, Cornell University, Ithaca, New York 14853, USA.\\
$^2$Theoretical Division, Los Alamos National Laboratory, Los Alamos, NM, 87545, USA.\\
$^3$Institute for Materials Science, Los Alamos, New Mexico 87545, USA.\\
$^4$Nordic Institute for Theoretical Physics (NORDITA), Roslagstullsbacken 23, S-106 91 Stockholm, Sweden}

\begin{abstract}

We study the effects of a finite density of negative-$U$ centers (NUCs) on the surface of a three-dimensional topological insulator. The surface Dirac fermions mediate a power-law interaction among the local Cooper pairs at the NUCs, and the interaction remains long-ranged for weak disorder. Superconductivity can be generated in the presence of a random distribution of NUCs. The NUCs play dual roles as both pair creator and pair breaker, and the competition of the two effects results in non-monotonic dependence of the mean field superconducting transition temperature on the density of NUCs. Global phase coherence is established through coupling the locally superconducting puddles via Josephson coupling. Rare fluctuations play important roles, and a globally superconducting phase can only be achieved at large enough concentration of NUCs. The $p$-wave component of the superconducting order parameter gives rise to frustration among the superconducting grains, which is captured by a Potts-XY type model. New phases with chiral order, glass order, and possibly topological order can then emerge in the system of superconducting grains.

\end{abstract}

\date{\today \ [file: \jobname]}

 \maketitle

\section{Introduction}

Topological phases of matter have recently attracted much attention in condensed matter physics. One prominent example is the topological insulator (TI), which is insulating in the bulk but possesses metallic surface states with linear dispersion [\onlinecite{Hasan10,Qi11}]. Such novel properties of TIs are protected by time reversal symmetry (TRS) and charge conservation symmetry (CCS). It is of both theoretical and practical importance to find ways to break these symmetries on the surface of TIs. Breaking TRS on the TI surface gives rise to topological magnetoelectric effect [\onlinecite{Qi08}], and breaking CCS leads to the formation of Majorana zero modes at the superconducting vortices [\onlinecite{Fu2008}]. In practice, there are two ways to break the surface symmetries. One way is to fabricate heterostructures of TIs and other symmetry broken materials, e.g. magnetic insulators that break TRS, superconductors that break CCS. Proximity effect then induces symmetry breaking at the TI surface. Another way is to deposite certain types of impurities on the TI surface, which has the advantage of simple experimental setup and better tunability for both bulk materials and thin films. Depositing magnetic impurities on the TI surface to break TRS has been extensively studied both theoretically and experimentally (see e.g. [\onlinecite{Liu09, Biswas10, Chen10, Schlenk13}]). Depositing impurities on the surface of TIs, or more generally Dirac materials including graphene, to generate pairing and break CCS was only proposed very recently by the present authors and collaborators [\onlinecite{She13, Fransson13}].

The basic idea of [\onlinecite{She13, Fransson13}] is to adsorb nonmagnetic molecules on the surface of Dirac materials, and use their vibration to produce local negative-$U$ interactions [\onlinecite{Anderson75}]. Local Cooper pairs can form at such negative-$U$ centers (NUCs), breaking CCS. Local electron density of states with a dilute concentration of NUCs has been studied in [\onlinecite{She13, Fransson13}], where it was found that strong enough coupling between electrons and local vibrations destroys the Dirac cone structure locally. In this paper, we study the collective behavior of a finite density of NUCs on the surface of TIs, with a focus on their superconducting properties. Here we treat the NUCs in a broader context. A NUC is generally understood as an impurity which forms an electronic state that prefers to be either empty or doubly occupied. It can have a phononic origin as considered in [\onlinecite{She13, Fransson13}], or it can have an excitonic origin, where certain valence state of an element is skipped (e.g. Tl$^{2+}$, Pb$^{3+}$, Sn$^{3+}$, Bi$^{4+}$, see e.g. [\onlinecite{Varma88}] ).

 Negative-$U$ superconductivity (see [\onlinecite{Micnas90}] and references therein) has been proposed as possible pairing mechanism for Pb- and K-doped BaBiO$_3$[\onlinecite{Varma88}], cuprates [\onlinecite{BarYam91}], Tl-doped PbTe [\onlinecite{Schmalian05}], and also as a generic mechanism to reduce phase fluctuations and enhance $T_c$ [\onlinecite{Kivelson08, Orgad12}]. A mean field theory (MFT) has been developed in [\onlinecite{Malshukov91, Malshukov92}] for a system of randomly distributed NUCs. We deviate from the previous approaches by considering NUCs coupled with massless Dirac fermions inherent to Dirac materials [\onlinecite{Vafek14, Wehling14}]. Furthermore, we study inhomogeneous superconductivity generated from rare fluctuations [\onlinecite{Balatsky97, Vojta0602, Spivak08, Nandkishore13}], which, to the best of our knowledge, has not been considered before for NUCs.

\section{pseudospin Kondo lattice model}

 We consider NUCs on the surface of a three dimensional TI. The local orbitals ($d_{i\alpha}$) on the NUCs hybridize with the Dirac fermions ($c_{{\bm k}\alpha}$) that propagate on the whole surface. We consider the onsite attractive interaction $U$ to be much larger than the hybridization amplitude. The singly-occupied states at the NUCs have much higher energy than the empty and doubly-occupied states, and can be projected out by the standard procedure of Schrieffer-Wolff transformation [\onlinecite{Schrieffer66}]. We consider the NUCs to be partially filled, and the empty sites and doubly occupied sites have the same energy. The local orbitals can then be represented by Anderson's pseudospins, with ${\cal T}^+_i=d^\dagger_{i\uparrow}d^\dagger_{i\downarrow}$, ${\cal T}^-_i=d_{i\downarrow}d_{i\uparrow}$, and ${\cal T}^z_i=\frac{1}{2}(n_i^d-1)$. These pseudospin operators obey SU(2) algebra as ordinary spins. The whole system is thus described by a pseudospin Kondo lattice model with Hamiltonian $H=H_c+H_K$, where 
\begin{eqnarray}
H_c&=&\sum_{\bm k\alpha\beta}c^{\dagger}_{{\bm k}\alpha}h_{\alpha\beta}({\bm k})c_{{\bm k}\beta},\\
H_K&=&\frac{J}{2N}\sum_i\left[
{\cal T}^+_ic_{i\downarrow}c_{i\uparrow}+h.c.+\frac{1}{2}\sum_\sigma {\cal T}^z_ic^\dagger_{i\sigma}c_{i\sigma}\right],
\label{Kondo}
\end{eqnarray}
with the kinetic term $h({\bm k})=v_F {\hat {\bm z}}\cdot ({\bm \sigma}\times{\bm k})$, the Fermi velocity $v_F$, Pauli matrices ${\bm \sigma}$, the unit vector ${\hat {\bm z}}$ perpendicular to the TI surface, and the charge Kondo coupling $J$.

\subsection{Single impurity: charge Kondo effect} 

We consider first the effect of coupling to Dirac fermions at a single NUC. With strong enough coupling, the pseudospins can be screened by the Dirac fermions, where they form pseudospin singlets with pairs of Dirac fermions, generating the charge Kondo effect. This effect has been studied for normal metals in [\onlinecite{Coleman91}], and can be easily generalized to the present case. The characteristic temperature scale, the charge Kondo temperature $T_K$, can be calculated from renormalization group or large-$N$ mean field theory, as for its spin counterpart [\onlinecite{Hewson}]. It is essentially determined by the local density of states (LDOS) $\rho({\bm R}, \varepsilon)$ at the impurity site (see e.g. [\onlinecite{Kotliar92}]). For Dirac fermions with a large Fermi surface, the LDOS is approximately constant, $\rho({\bm R}, \varepsilon)\simeq \rho_0$, and $T_K$ is of the usual Fermi-liquid form $T_K\simeq \epsilon_F \exp(-1/J\rho_0)$,
 with Fermi energy $\epsilon_F $ [\onlinecite{Coleman91}]. For Dirac fermions with $\mu=0$, the LDOS is linear in energy, $\rho({\bm R}, \varepsilon)= |\varepsilon|/2\pi v_F^2$, and one has $T_K\simeq \Lambda(1-J_c/J)$, with a cutoff $\Lambda$. In this case, the charge Kondo temperature is only nonvanishing when the Kondo coupling is larger than the critical value $J_c=2\pi v_F^2/\Lambda$. Otherwise $T_K=0$. With a linear density of states, the Dirac fermions are much less effective in screening the NUCs than a normal electron gas.

\subsection{Two impurities: pseudospin interactions}

 We consider next the interaction between two NUCs mediated by the Dirac fermions, 
\begin{equation}
H_{ij}= I_{ij}^{\perp} {\cal T}^z_i{\cal T}^z_j+I^{\parallel}_{ij}\left( {\cal T}^x_i{\cal T}^x_j+{\cal T}^y_i{\cal T}^y_j\right).
\end{equation}
 The couplings are determined by fermion bubbles in the charge and pairing channels [\onlinecite{Malshukov91, Malshukov92}],
\begin{eqnarray}
I^{\perp}_{ij}&=&\frac{J^2T}{2N^2}\sum_n {\rm Tr} \left[{\cal G}_{ij}(\omega_n){\cal G}_{ji}(\omega_n)\right],\nonumber\\
I^{\parallel}_{ij}&=&\frac{J^2T}{2N^2}\sum_n {\rm Tr} \left[\sigma_y{\cal G}_{ij}^T(\omega_n) \sigma_y{\cal G}_{ij}(-\omega_n)\right].
\end{eqnarray}

Consider first the clean case. For $\mu$ large, the Green's function has the same asymptotic form as that of a two-dimensional electron gas, i.e. 
\begin{equation}
{\cal G}_{ij}(\omega_n)\sim r^{-1/2} \exp[-\frac{|\omega_n|r}{v_F}+i{\rm sgn}(\omega_n)k_Fr].
\end{equation}
Then one obtains at zero temperature,
\begin{equation}
I^{\perp}_{ij}\sim \cos(2k_Fr)/r^2
\end{equation}
with $2k_F$-oscillation, and
\begin{equation}
I^{\parallel}_{ij}\sim 1/r^2,
\end{equation}
decaying monotonically. For $\mu=0$, the Green's function reads 
\begin{equation}{\cal G}_{ij}(\omega_n)=\frac{-i\omega_n}{2\pi v_F^2}K_0\left(\frac{|\omega_n|r}{v_F} \right)+{\hat z}\cdot({\hat r}\times{\bm \sigma})\frac{i|\omega_n|}{2\pi v_F^2}K_1\left(\frac{|\omega_n|r}{v_F} \right),
\end{equation}
 where $K_{\alpha}(x)$ are the modified Bessel functions of the second kind. The pseudospin interactions are thus of the form 
\begin{equation}
I^{\perp}({\bm r})\sim I^{\parallel}({\bm r})\sim\int d\omega\omega^2K^2_\alpha(|\omega|r/v_F)\sim 1/r^3,
\label{RKKY}
\end{equation}
 decaying faster than the case with large chemical potential.  At finite temperatures, the interactions have an exponential decay $I_{ij}\sim e^{-r/l_T}$, controlled by the thermal length $l_T\sim v_F/T$. We note that the interaction between two Cooper pairs does not oscillate with their separation, in contrast to the usual RKKY interaction. Here the interaction is always ferromagnetic, tending to align the phases of Cooper pairs.

 In the presence of weak disorder, the Green's function acquires a random phase shift [\onlinecite{Spivak86, Bulaevskii86}], with 
\begin{equation}
{\cal G}_{ij}(\omega_n)\to {\cal G}_{ij}(\omega_n)\exp[{-i\frac{{\rm sgn}(\omega_n)}{v_F}}\int ds  {\tilde V}({\bm r})],
\end{equation}
 where ${\tilde V}({\bm r})$ is the impurity potential, and the integral is over the straight line connecting ${\bm r}_i$ and ${\bm r}_j$. For large $\mu$, the charge part then becomes 
\begin{equation}
I^{\perp}_{ij}\sim \cos(2k_Fr+\varphi_{ij})/r^2,
\end{equation}
 with a random phase shift $\varphi_{ij}$. Averaging over the impurity configurations gives 
\begin{equation}
\langle I^{\perp}_{ij}\rangle\sim r_{ij}^{-2}\cos(2k_Fr)\exp(-r_{ij}/l_e),
\end{equation}
 which is exponentially suppressed outside the mean free path $l_e$. However, in the pairing channel, the random phase shift from the two electrons that form a Cooper pair cancels, in accordance with Anderson's theorem. This gives the remarkable result that $I^{\parallel}_{ij}$ remains of the long-range power-law form even after taking the impurity average, 
\begin{equation}
\langle I^{\parallel}_{ij}\rangle\sim 1/r^2.
\label{Ipara}
\end{equation}
This result is very different from its spin counterpart, where the impurity averaged RKKY interaction decays exponentially, while the even moments of the interactions remain long ranged, signaling large amplitude fluctuations [\onlinecite{Spivak86, Bulaevskii86, Jagannathan88}]. We note that for $\mu=0$, due to the generation of a finite density of states by disorder for the surface Dirac fermions [\onlinecite{Shon98}], the power in $I^{\parallel}_{ij}$ remains the same as the case for large $\mu$. Thus disorder enhances the coupling among the NUCs with large separations.

\begin{figure}
\begin{centering}
\includegraphics[width=0.31\linewidth]{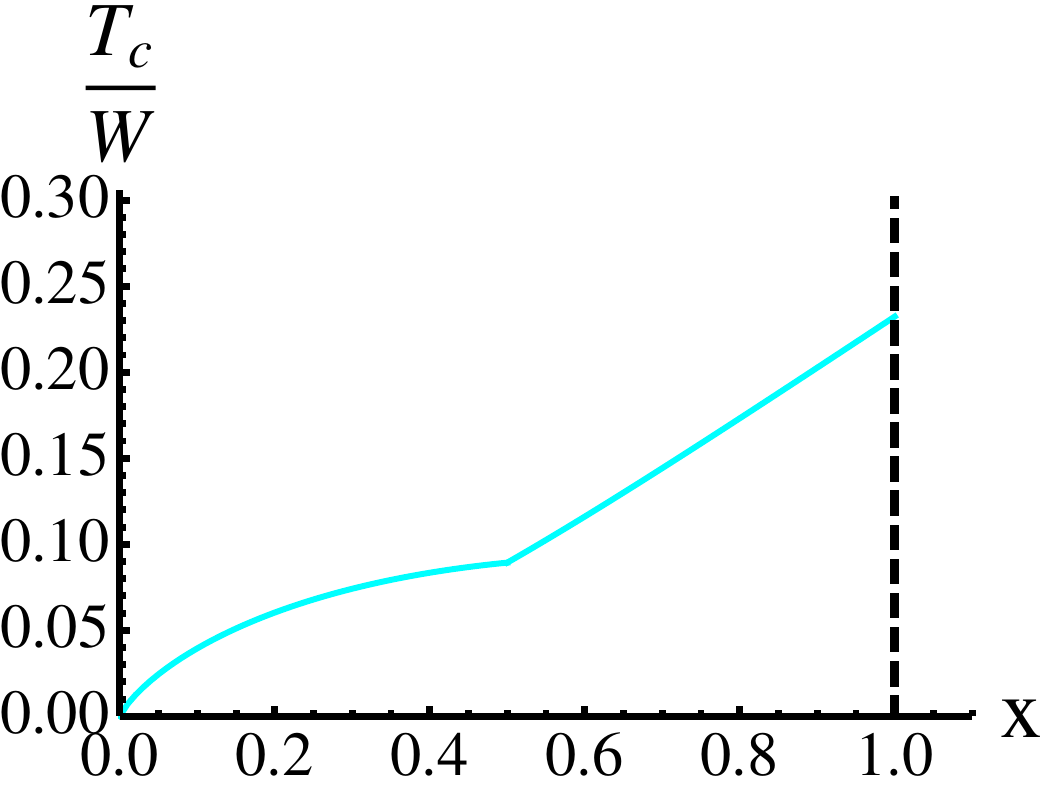} 
\includegraphics[width=0.31\linewidth]{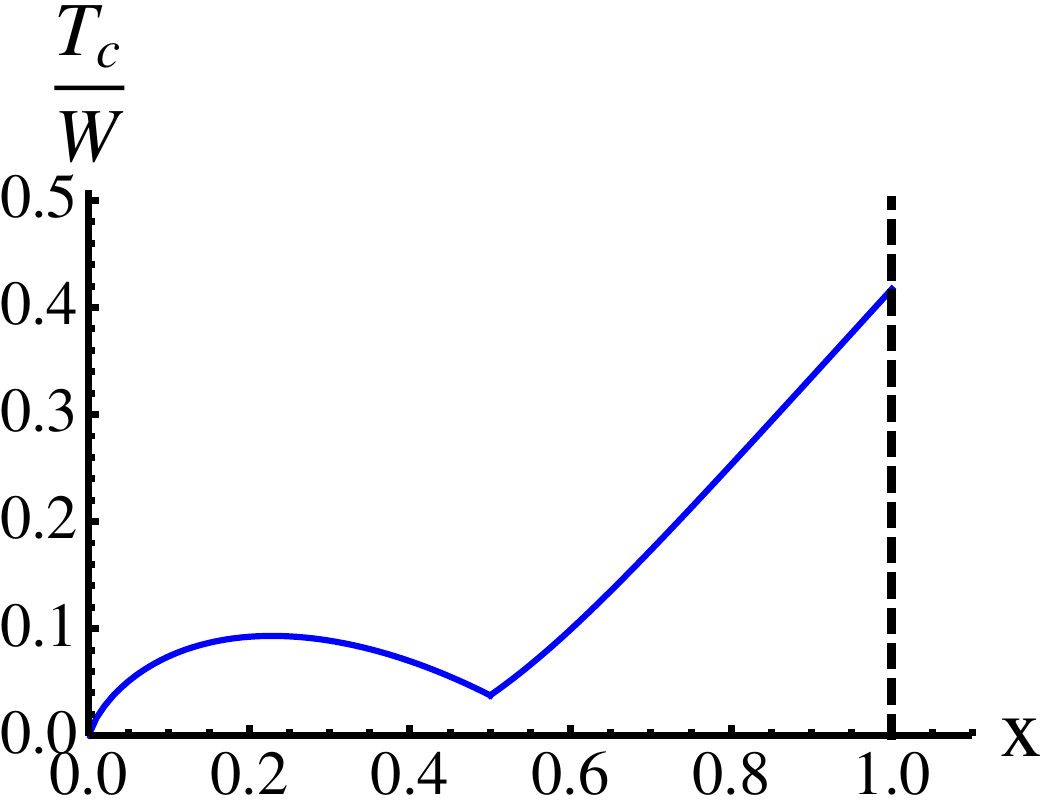} 
\includegraphics[width=0.31\linewidth]{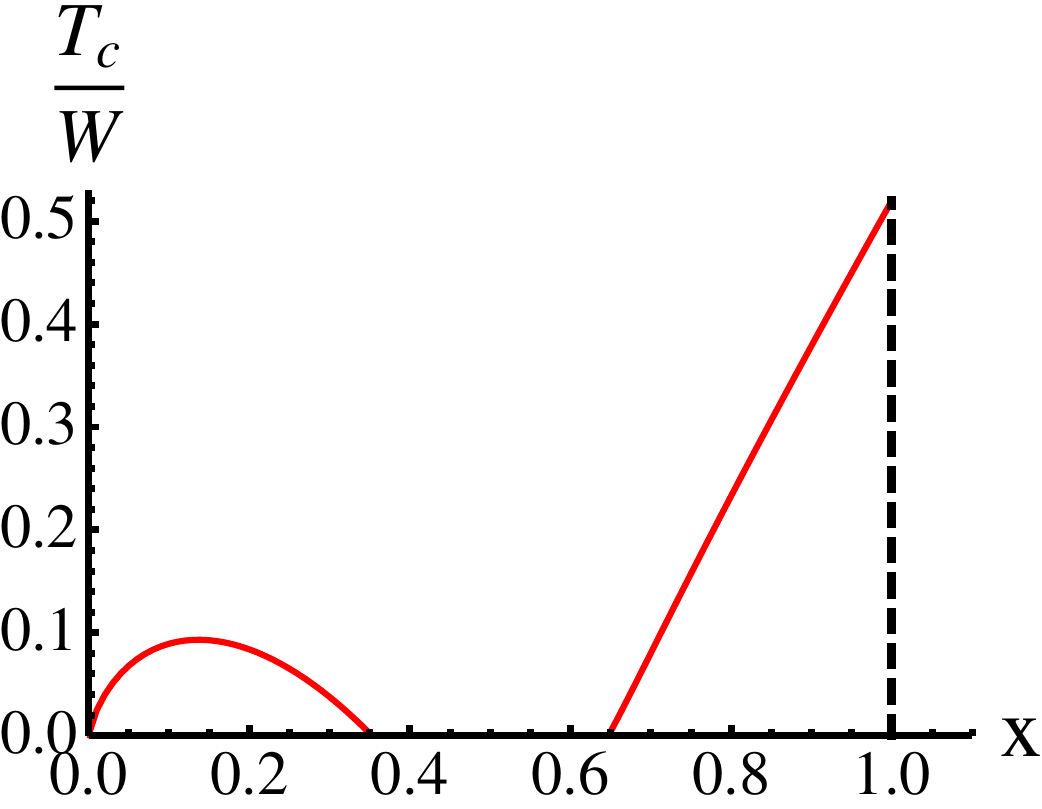} 
\end{centering}
\caption{Superconducting transition temperature as function of the concentration of  negative-$U$ centers for Dirac fermions with large Fermi surface. Here $\pi J^2 N(0)/8=5, 15, 25$ (from left to right). We have used $1/2\tau_e=\pi x J^2N(0)/8$ for $x<1/2$, and $1/2\tau_e=\pi (1-x) J^2N(0)/8$ for $x>1/2$.}
\label{Tc-metal}
\end{figure}

\begin{figure}
\begin{centering}
\includegraphics[width=0.31\linewidth]{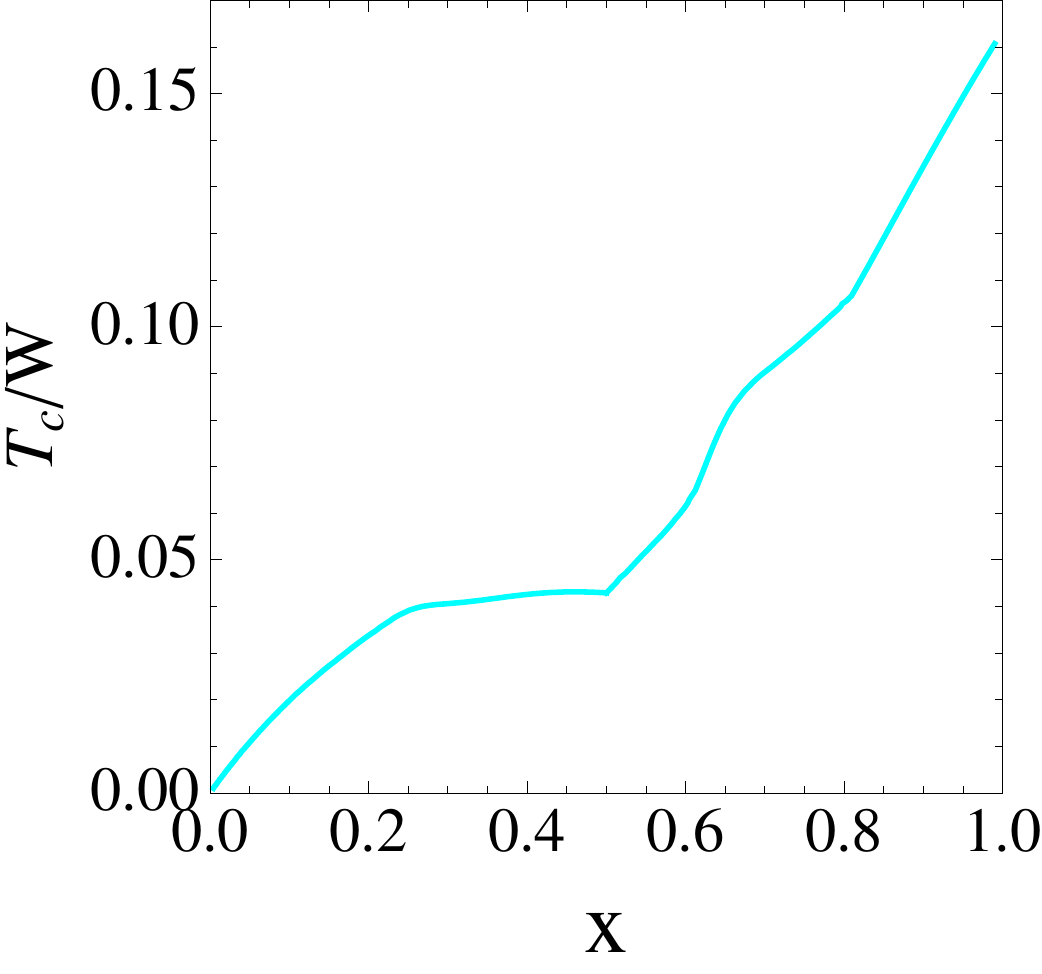} 
\includegraphics[width=0.31\linewidth]{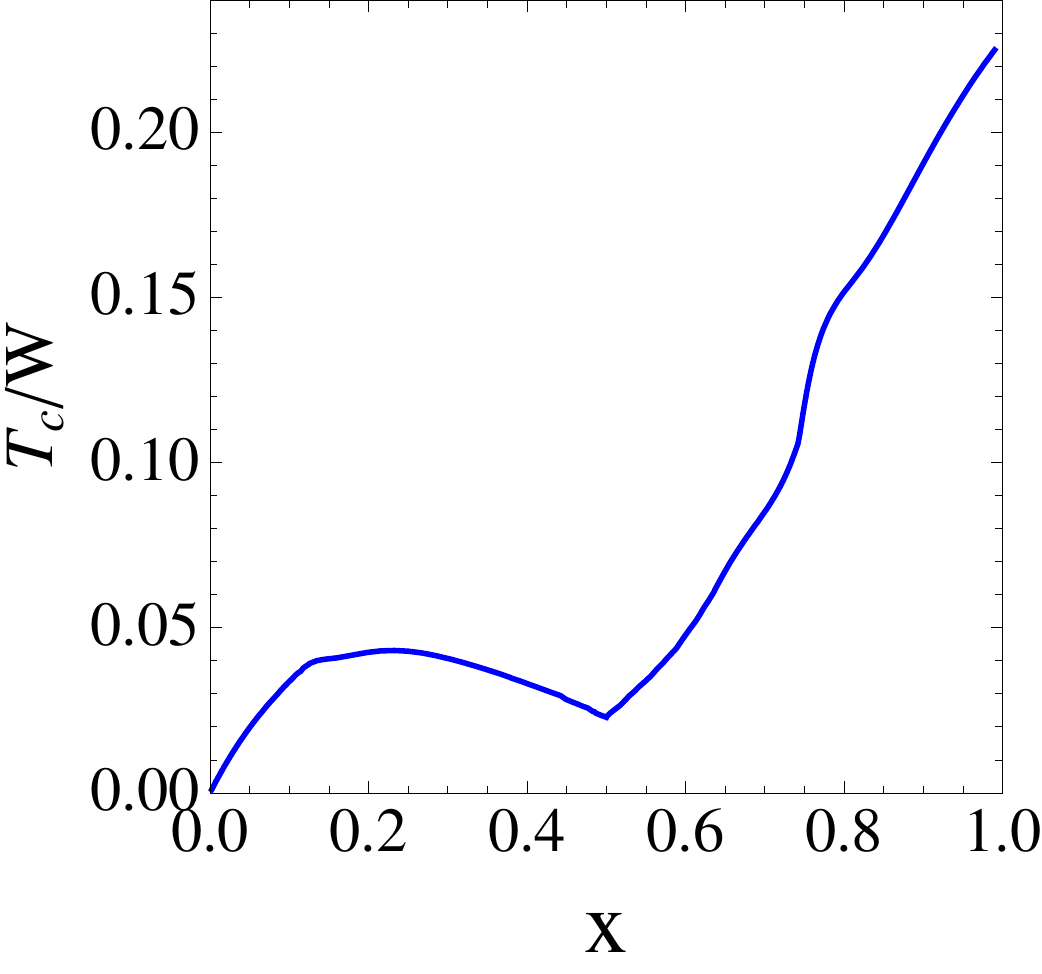} 
\includegraphics[width=0.31\linewidth]{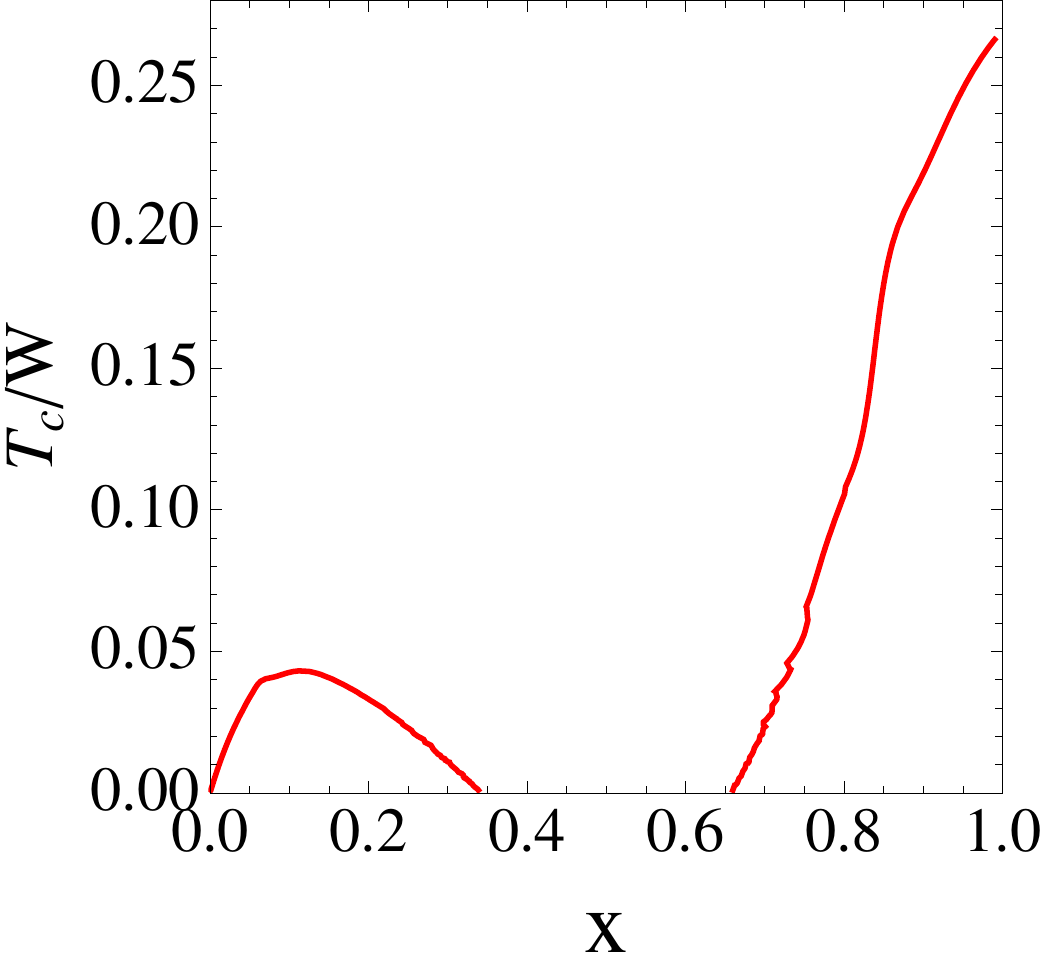} 
\end{centering}
\caption{The same plot as Fig.~\ref{Tc-metal} for Dirac fermions with $\mu=0$, and $J^2/(8\pi v_F^2)=1.5, 3, 6$ (from left to right). }
\label{Tc-Dirac}
\end{figure}

\subsection{Many impurities: superconductivity from MFT} 

 A finite density of NUCs naturally leads to superconductivity. To get superconductivity, both pairing and phase coherence need to be achieved. We study first the onset of pairing in MFT [\onlinecite{Malshukov91, Malshukov92}], and then proceed to consider the transition to a globally phase coherent state using a rare-fluctuation based approach. Condensation of the local Cooper pairs at the NUCs induces a pairing interaction among the Dirac fermions, 
\begin{equation}
\delta H_c=\frac{J}{2N}\sum_i( c^\dagger_{i\uparrow}c^\dagger_{i\downarrow}\langle{\cal T}^-_i\rangle+c_{i\downarrow}c_{i\uparrow}\langle{\cal T}^+_i\rangle).
\end{equation}
 A pairing gap $\Delta^*_i=\langle c^\dagger_{i\uparrow}c^\dagger_{i\downarrow} \rangle$ is thus generated for the Dirac fermions, whose value can be determined from
\begin{equation}
 \Delta^*_i=-\frac{J}{2}\sum_j\int_0^\beta d\tau \langle c_{j\downarrow}c_{j\uparrow}(\tau)c^\dagger_{i\uparrow}c^\dagger_{i\downarrow}(0)\rangle  \langle {\cal T}^+_j \rangle.
\end{equation}
 The pairing gap of Dirac fermions acts as a potential well for the local Cooper pairs, with Zeeman type coupling 
\begin{equation}
H_s=\frac{J}{2N}\sum_i \left(\Delta^*_i {\cal T}^-_i+\Delta_i  {\cal T}^+_i\right),
\end{equation}
 from which one obtains 
\begin{equation}
\langle{\cal T}^+_i\rangle={\rm Tr}[ {\cal T}^+_i \exp( -\beta H_s)]/{\rm Tr}[ \exp( -\beta H_s)].
\end{equation}
 To leading order in the coupling strength, we have 
\begin{equation}
\langle{\cal T}^+_i\rangle\sim\beta  J\Delta_i^*.
\end{equation}
 Substituting it back to the gap equation, we obtain the mean field $T_c$ equation
\begin{equation}
1-\frac{xJ^2}{16}\sum_{{\bm k}n}{\rm Tr} \left[\sigma_y{\hat{\cal G}}^T({\bm k}, \omega_n)\sigma_y{\hat{\cal G}}(-{\bm k}, -\omega_n) \right]=0,
\end{equation}
with $x$ the concentration of NUCs. Here the disorder-averaged Dirac fermion Green's function ${\hat{\cal G}}({\bm k}, \omega_n)$ is a $2\times 2$ matrix in spin space. This equation is of the RPA form typical for a Stoner-type instability, $1-g\chi_{\rm pair}=0$, with an effective coupling $g\sim xJ^2$, and the pair susceptibility $\chi_{\rm pair}\sim{\rm Tr}[\sigma_y{\hat{\cal G}}^T\sigma_y{\hat{\cal G}}]$. It is a direct generalization of the results of [\onlinecite{Malshukov91, Malshukov92}] to spin-orbit coupled systems.

For Dirac fermions with large chemical potential, one can consider only the conduction band. In the Born approximation, the disorder-averaged Green's function reads 
\begin{equation}
{\cal G}({\bm k}, \omega_n)=\frac{1}{2}\frac{\sigma^0+{\bm \sigma}\cdot {\bm e}_{\bm k}}{i{\tilde \omega}_n-\xi_{\bm k}},
\end{equation}
 with kinetic energy $\xi_{\bm k}=v_Fk-\mu$, and unit vector ${\bm e}_{\bm k}={\bm k}/k=(\cos\theta_{\bm k}, \sin\theta_{\bm k}, 0)$ (see e.g. [\onlinecite{Tkachov11}]). The frequency dependence is renormalized by pseudospin flip scattering, with 
\begin{equation}
{\tilde \omega}_n=\omega_n+\frac{1}{2\tau_e}{\rm sgn}(\omega_n).
\end{equation}
 For dilute NUCs, the scattering rate is 
\begin{equation}
1/2\tau_e=\pi x J^2N(0)/8,
\end{equation}
 where $N(0)$ is the conduction electron density of states at the Fermi level. For dense NUCs, the impurities are the vacancies, and one has
\begin{equation}
1/2\tau_e=\pi (1-x) J^2N(0)/8.
\end{equation}
 The resulting gap equation is the same as that of a normal metal [\onlinecite{Malshukov91, Malshukov92}], i.e. 
\begin{equation}
(2\tau_p)^{-1}\sum_{n}(|\omega_n|+1/2\tau_e)^{-1}=1,
\end{equation}
 where $(2\tau_p)^{-1}=\pi x J^2N(0)/8$ characterizes the pairing strength. We have $\tau_p=\tau_e$ for dilute impurities. Carrying out the frequency summation, one obtains [\onlinecite{Malshukov91, Malshukov92}]
\begin{equation}
2\pi\tau_pT_c= \Psi\left(\frac{1}{2}\right)- \Psi\left(\frac{1}{2}+\frac{1}{4\pi\tau_e T_c}\right)+\ln\frac{W}{T_c},
\end{equation}
where $\Psi$ is the digamma function, and $W$ is a cutoff. 

One can see that NUCs play dual roles for superconductivity: they produce attactive interactions that drive pairing ($\tau_p$ term), and the randomness of their positions leads to pair breaking effects ($\tau_e$ term) as in the case of magnetic impurities [\onlinecite{BalatskyRMP}]. The competition of the two effects is manifest for dilute impurities. One can see from the numerical solution of the gap equation (Fig.~\ref{Tc-metal}) non-monotonic behavior of $T_c$ for low impurity concentrations. With increasing pairing strength, superconductivity is generated, and $T_c$ first increases with the concentration of NUCs. At higher concentrations, the pair breaking effect takes over, and $T_c$ may be suppressed. For concentrations close to unity, as $x$ increases, impurity scattering gets weaker, while pairing gets stronger. Thus $T_c$ increases monotonically (see Fig.~\ref{Tc-metal}).

For Dirac fermions with $\mu=0$, the linear dispersion leads to a different form of gap equation. Consider the Green's function of the form ${\hat{\cal G}}^{-1}({\bm k}, \omega_n)= i{\tilde\omega}_n-{\hat {\cal H}}_c$, with the Hamiltonian ${\hat {\cal H}}_c=v_F(\sigma_x k_y-\sigma_y k_x)$, and the self-energy corrections incorporated in ${\tilde\omega}_n$. The gap equation becomes
\begin{equation}
1-\frac{\alpha_p}{2}\sum_n\ln\frac{W}{|{\tilde\omega}_n|}=0,
\end{equation}
with the pairing strength $\alpha_p\simeq xJ^2/(8\pi v_F^2)$. The effect of impurity scattering in this case is more subtle than the case with large Fermi surface. A finite density of states is generated at the Dirac point by impurity scattering, for which simple Born approximation is not enough. The essential physics is captured by the self-consistent Born approximation (SCBA) [\onlinecite{Shon98}], and the result takes the form 
\begin{equation}
{\tilde\omega}_n= \omega_n( 1+\frac{1}{\alpha})+\Gamma_0{\rm sgn}(\omega_n),
\end{equation}
 for $|\omega_n| \ll \Gamma_0$, and 
\begin{equation}
{\tilde\omega}_n= \omega_n[ 1+\alpha\ln(W/|\omega_n|)],
\end{equation}
 for $|\omega_n| \gg \Gamma_0$ [\onlinecite{Mirlin06}]. A new energy scale $\Gamma_0=W e^{-1/\alpha}$ is generated, with the dimensionless scattering rate $\alpha\simeq xJ^2/(8\pi v_F^2)$ for dilute impurities, and $\alpha\simeq (1-x)J^2/(8\pi v_F^2)$ for dense impurities. Impurity scattering gives rise to a finite density of states $\rho_0\sim \Gamma_0/(v_F^2\alpha)$ at the Dirac point. With the knowledge of the self-energy corrections, the gap equation can be solved numerically, and the result is qualitatively the same as the case with a large Fermi surface (see Fig.~\ref{Tc-Dirac}).

 This result has important consequencies for the competition between superconducting ordering and charge Kondo effect. In the weak coupling region, due to the generation of a finite density of states, the charge Kondo temperature is of the Fermi liquid exponential form, and superconductivity dominates. At large coupling, the superconducting $T_c$ saturates or even decreases, and the charge Kondo effect dominates. Near $T_K\sim T_c$, the competition of the two effects can give rise to a reentrance to normal state at low temperatures [\onlinecite{Schmalian05}].

\section{Superconductivity from rare fluctuations} 

\begin{figure}
\begin{centering}
\includegraphics[width=0.8\linewidth]{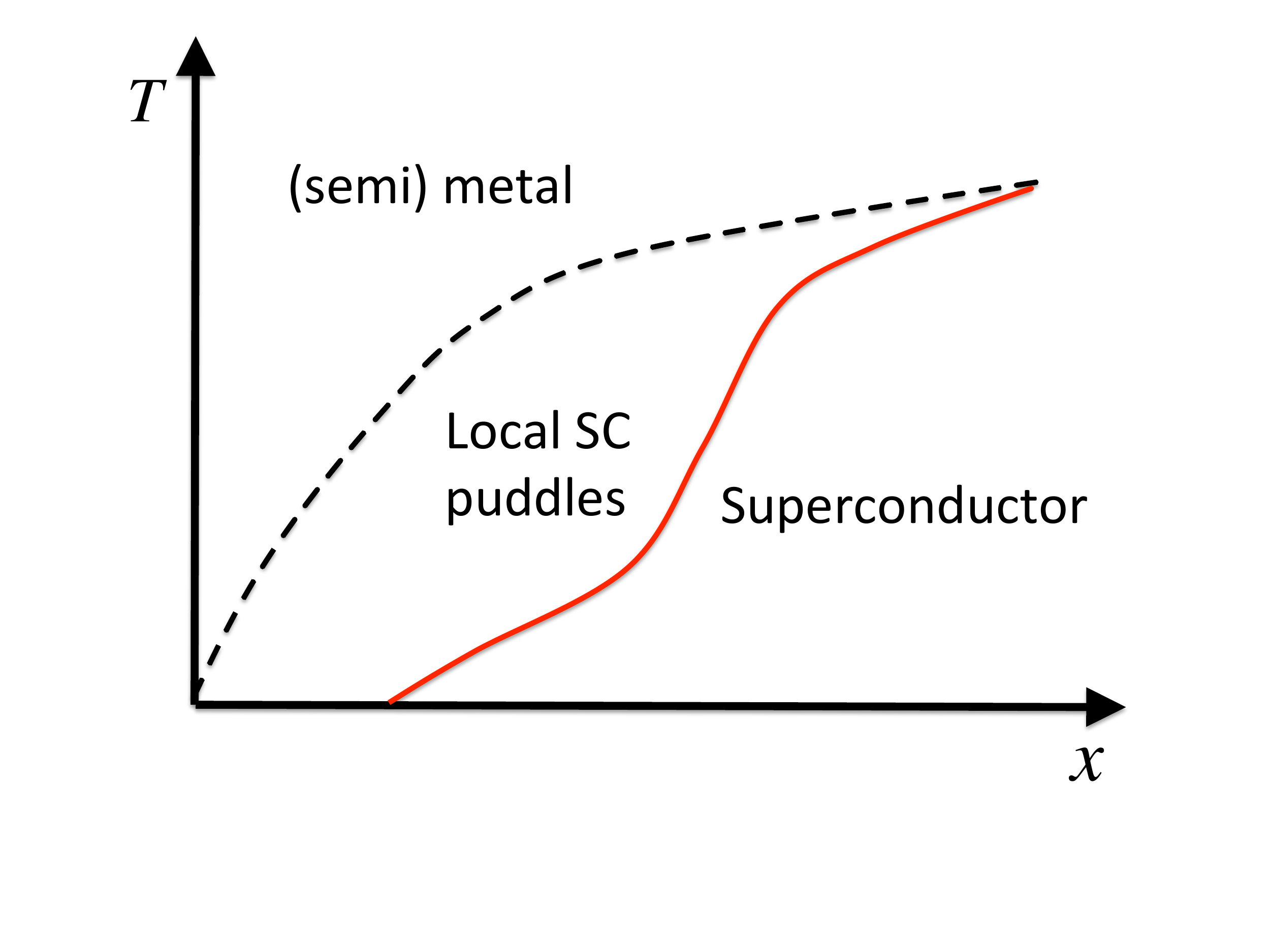} 
\end{centering}
\caption{Schematic phase diagram for rare fluctuation generated superconductivity. The dashed line represents a crossover from a high temperature (semi)-metallic phase to a phase with locally superconducting puddles. The solid line represents a phase transition into the globally superconducting phase, in which global phase coherence is established among the local superconducting puddles via Josephson coupling.}
\label{puddle}
\end{figure}

Now we go beyond MFT, and consider inhomogeneous superconductivity from rare fluctuations (see [\onlinecite{Balatsky97, Vojta0602, Spivak08, Nandkishore13}] and references therein).
For a finite density of randomly distributed NUCs, there will be rare regions with dense NUCs devoid of vacancies. These regions will be superconducting locally at a much higher transition temperature $T_c^0$ than the global superconducting $T_c$. When the Josephson coupling between these superconducting puddles is strong enough, phase coherence can be achieved, and the whole TI surface will enter the superconducting phase (see Fig.~\ref{puddle}).

When the concentration $x$ is larger than the percolation threshold $x_p$ (e.g. $x_p\simeq 0.59$ on a square lattice), a large superconducting cluster is formed, ensuring phase coherence. In this case, the superconducting transition temperature is  constrained by the mean field transition temperature. We note that near the percolation threshold, since the superconducting cluster is fractal-like, the mean field transition temperature is suppressed by impurity scattering. We consider below the case of dilute concentration of NUCs with $x\ll x_p$.

For a given concentration $x$, the probability to find a region of size $R$ devoid of NUC vacancies is $w(R)\sim x^{(R/a)^2}\sim \exp[-p(R/a)^2]$,
 with lattice spacing $a$, and $p\sim -\ln x>0$. Thus smaller sized puddles are exponentially likely to occur. For these puddles to be superconducting locally, the size of the puddles needs to be larger than the local coherence length $\xi\sim v_F/T_c^0$. So the optimum size of the superconducting puddles is $R\sim \xi$.

The Josephson coupling between two superconducting puddles with puddle size $R$ and interpuddle spacing $L$, with $L\gg R$, is of the form $J_p\sim \frac{v_F R^2}{L^3}e^{-L/l_T}$,
 when mediated by Dirac fermions with $\mu=0$ [\onlinecite{Gonzalez08}]. The $1/L^3$ dependence can be deduced from Eq.(\ref{RKKY}). Disorder generates finite density of states even for $\mu=0$, and changes the power. But here we still use this result as a lower bound. The interpuddle spacing $L$ needs to be smaller than the thermal length $l_T$ to get an appreciable Josephson coupling. When this condition is satisfied, i.e. $L<l_T$, we have approximately $J_p\sim \frac{v_F R^2}{L^3}$.

For a dilute concentration of NUCs, the probability that a given region of size $R\sim \xi$ is in a superconducting phase is $P(\xi)\sim \int_\xi^\infty dR w(R)\sim {\rm erfc}[\sqrt{p}(\xi/a)]\sim e^{-p(\xi/a)^2}$, for $\xi\gg a$.
 The typical interpuddle spacing is then 
\begin{equation}
L\sim \xi P^{-1/2}(\xi)\sim\xi \exp[\frac{1}{2}p(\xi/a)^2],
\end{equation}
 which is much larger than the typical puddle size. The condition $L<l_T$ gives the constraint 
\begin{equation}
T_c\leq T_c^0\exp[-\frac{1}{2}p(\xi/a)^2].
\end{equation}
 The Berezinsky-Kosterlitz-Thouless transition temperature is obtained from $T_{\rm BKT}\sim J_p$, which then determines the superconducting transition temperature for small $x$,
\begin{equation}
T_c\sim T_{\rm BKT}\sim T_c^0 \exp\left[ -\frac{3}{2}p\left( \frac{\xi}{a} \right)^2\right]\sim T_c^0 x^{\frac{3}{2}(\xi/a)^2}.
\end{equation}
For $\xi\sim v_F/T_c^0\gg a$, $T_c$ essentially vanishes at small $x$. As emphasized in [\onlinecite{Spivak08}], the interaction between the superconducting phase fluctuations and the quantum fluctuations of the electromagnetic field further suppresses coherence of each superconducting puddle, giving rise to a quantum phase transition at finite $x$. Thus for a dilute concentration of NUCs, although the mean field transition temperature can be appreciable, there are only local superconducting puddles, and the whole surface is not superconducting due to lack of global phase coherence (see Fig.~\ref{puddle}). 

\section{Frustration in superconducting grains}
 
The NUC-based setup has more local tunability than the proximity induced superconductivity. As an application of negative-$U$ superconductivity, we consider here possible new phases generated by an ensemble of superconducting grains, i.e. Josephson junction arrays, on the surface of TI. Without frustration, the phases of the superconducting grains will order ferromagnetically, and hence superconductivity is the only possible order. The presence of frustration is associated with the breaking of TRS. For $s$-wave superconductors, external magnetic field or magnetic impurities are needed to break TRS and generate frustration. A remarkable property of unconventional superconductors is that TRS can be spontaneously broken in a superconducting grain, even if the corresponding bulk phase is time reversal invariant [\onlinecite{Sigrist95, Spivak14}]. Hence frustration can be generated in such grains of unconventional superconductors.

On the surface of TI with finite chemical potential, due to strong spin-orbit coupling, superconductivity is a mixture of $s$- and $p$-wave. The $p$-wave component leads to frustrated interactions. Consider large chemical potential, in the helicity basis, after projecting to a single helicity, we have effectively a one band model of spinless fermions, with $p_x\pm i p_y$-pairing inside the superconducting puddles. The gap at puddle $i$ can be written as $\Delta_i=e^{i\phi_i}\Delta^{(i)}_p  \eta_{ia}p_a$, where $\phi_i$ is the phase of the order parameter, $\Delta^{(i)}_p$ the amplitude, and $(\eta_{ix}, \eta_{iy})=\frac{1}{\sqrt{2}}(\pm 1, \pm i)$ represents the orbital orientation. The corresponding Josephson coupling is (see e.g. [\onlinecite{Spivak14}]) 
\begin{equation}
H_J^{(p)}=-\sum_{i\neq j}{\cal A}_{ij}{\rm Re}\left[ e^{i(\phi_i-\phi_j)}  \eta_{ia}\eta^*_{jb} \frac{\partial}{\partial r_i^a} \frac{\partial}{\partial r_j^b}\frac{1}{|{\bm r}_i-{\bm r}_j|^2}\right],
\end{equation} 
where ${\cal A}_{ij}\sim\nu \Delta^{(i)}_p \Delta^{(j)}_p$, with $\nu$ the density of states at the Fermi level. With $ \frac{\partial}{\partial r_i^a} \frac{\partial}{\partial r_j^b}\frac{1}{|{\bm r}_i-{\bm r}_j|^2}=2\frac{\delta_{ab}-4{\hat r}_a{\hat r}_b}{r^4}$, we have 
\begin{equation}
H_J^{(p)}=-\sum_{i\neq j}   J^{(p)}_{ij} {\rm Re}\left[ e^{i(\phi_i-\phi_j)} \eta_{ia}\eta^*_{jb}  (\delta_{ab}-4{\hat r}_a{\hat r}_b)\right],
\end{equation}
where $J^{(p)}_{ij}\sim \nu \Delta^{(i)}_p \Delta^{(j)}_p/r^4$. One can see that the coupling is strongly orientation dependent.

The orientation can be parameterized as $\eta_{ix}+\eta_{iy}=\exp\left[i 2\pi f \left( \frac{1}{2}+n_i\right)\right]$, with $f=1/4$ and $n_i=0, 1, 2, 3$. Hence on each puddle, in addition to the U(1) phase $\phi_i$, one has also local discrete degrees of freedom described by a four-state Potts type model. The coupling then can be written as
\begin{equation}
H_J^{(p)}=-\sum_{i\neq j}   J^{(p)}_{ij} \left[ C_{ij}\cos(\phi_i-\phi_j) +S_{ij}\sin(\phi_i-\phi_j)\right],
\end{equation}
with the Potts part of the Hamiltonian
\begin{eqnarray}
C_{ij}&=&-\cos\frac{\pi}{2}(n_i-n_j)+2\cos 2\theta_{\bm r}\sin\frac{\pi}{2}(n_i+n_j),\nonumber\\
S_{ij}&=&2\sin 2\theta_{\bm r}\sin\frac{\pi}{2}(n_i-n_j),
\end{eqnarray}
where $\theta_{\bm r}$ is the angle of the vector connecting two puddles.

The $s$-wave component of the order parameter has the usual Josephson coupling
\begin{eqnarray}
H_J^{(s)}=-\sum_{i\neq j} J^{(s)}_{ij}\cos\left(\phi_i-\phi_j\right),
\end{eqnarray}
with $J^{(s)}_{ij}\sim 1/r^2$, decaying much slower than the $p$-wave component. For finite chemical potential, both components are present (see e.g. [\onlinecite{Nandkishore13}]). At large interpuddle spacing, the Josephson coupling is dominated by the $s$-wave component. As realized in [\onlinecite{Spivak08, Spivak14}], when the interpuddle spacing is much larger than the puddle size, the grains of unconventional superconductors behave as a $s$-wave superconductor at large scales. 

When the interpuddle spacing becomes comparable to the puddle size, the coupling arising from the $p$-wave component is appreciable, and the interactions among the puddles are frustrated. Such frustrated $XY$-models have a rich phase diagram. We consider several limiting cases in the following.

Consider for example arranging the puddles to form a triangular lattice, and tuning the order parameters to have the same orientation at each puddle, when the $p$-wave component dominantes the coupling, one obtains an antiferromagnetic XY model on a triangular lattice (i.e. $S_{ij}=0$, $C_{ij}=-1$ for $n_i=n_j$). This model is in the same universality class as the fully frustrated XY model [\onlinecite{Teitel8302}], the phase diagram of which has been extensively studied (see [\onlinecite{Korshunov06, Vicari05}] and references therein). In addition to the $U(1)$ symmetry, the Hamiltonian is invariant under the global $Z_2$ symmetry: $\phi_i\to -\phi_i$. Hence domain walls can appear in the system, and the system can exhibit a chiral phase where the $Z_2$ symmetry is broken, while the $U(1)$ symmetry is preserved. 

Another example is to have the location of the puddles and the orientation of the order parameter at each puddle to be random. Their effect can be modeled by a random gauge field $A_{ij}$ at each bond, and the Haniltonian is of the form of a random phase XY model
\begin{equation}
H_J=-\sum_{i\neq j}   {\tilde J}_{ij} \cos(\phi_i-\phi_j-A_{ij}).
\end{equation} 
Here the coupling strength is also random, but its effect is less important as compared to the phase part, which gives rise to frustration. When the random phase disorder is strong enough, the quasi-long-range-order of the XY model is destroyed, and a glass phase is expected (see [\onlinecite{Korshunov06, Vicari10}] and references therein).

Finally we would like to mention that the system of superconducting grains on the surface of TI  provides a natural setup to realize more exotic phases with surface topological orders [\onlinecite{Metlitski13, Bonderson13, Senthil13, Vishwanath13}]. These phases preserve both TRS and CCS. However, in order to obtain new phases beyond the original superconducting state, frustration is required, and hence the breaking of TRS. These constitute an apparent paradox, which can be solved in the setup of coupled grains in two steps. In the grain system, TRS is first broken for a particular configuration of the order parameter orientations $\{n_i\}$, which generates frustration among the phase variables. Frustration leads to the condensation of vortex bound states [\onlinecite{Metlitski13, Bonderson13, Senthil13, Vishwanath13}], and the phase variables can be driven to a liquid state. Then fluctuations of the order parameter orientations restore TRS, i.e. TRS is preserved after summing over different realizations of the order parameter orientations, $\int{\cal D}n_i\int{\cal D}\phi_i\exp\left( -\beta H_J[n_i, \phi_i]\right)$. Further investigations are needed to establish an explicit connection between the above Potts-XY type model and the surface topological orders.

\section{Conclusions} 

We have shown in this paper that by depositing a finite density of randomly distributed NUCs on the surface of TI, a superconducting surface termination of TI can be achieved. To generate superconductivity, both pairing and phase coherence are required. In the mean field approach, we have shown that NUCs play dual roles for pairing, as both pair creator and pair breaker, which results in nonmonotonic dependence of the mean field superconducting transition temperature on the concentration of NUCs. In the puddle based approach, local superconducting puddles are first created, which then interact via Josephson coupling to establish global phase coherence. The concentration of NUCs needs to exceed certain threshold to generate a globally superconducting phase. New phases can be generated by incorporating frustration among the superconducting grains. The NUCs provide a new fundamental element for engineering functional Dirac materials.

We acknowledge valuable discussions with Annica M. Black-Schaffer, Chih-Chun Chien, Jonas Fransson, Matthias Graf, Christopher L. Henley, Dmytro Pesin, Jie Ren, Boris Spivak, Eddy Timmermans, Abolhassan Vaezi, C. -C. Joseph Wang and Jian-Xin Zhu. Work was supported by VR 2012-2983, ERC DM-321031 and US DOE. Work at Cornell was supported by the Cornell Center for Materials Research with funding from the NSF MRSEC program (DMR-1120296).

\bibliographystyle{apsrev}
\bibliography{strings,refs}

\begin{thebibliography}{47}
\expandafter\ifx\csname natexlab\endcsname\relax\def\natexlab#1{#1}\fi
\expandafter\ifx\csname bibnamefont\endcsname\relax
  \def\bibnamefont#1{#1}\fi
\expandafter\ifx\csname bibfnamefont\endcsname\relax
  \def\bibfnamefont#1{#1}\fi
\expandafter\ifx\csname citenamefont\endcsname\relax
  \def\citenamefont#1{#1}\fi
\expandafter\ifx\csname url\endcsname\relax
  \def\url#1{\texttt{#1}}\fi
\expandafter\ifx\csname urlprefix\endcsname\relax\def\urlprefix{URL }\fi
\providecommand{\bibinfo}[2]{#2}
\providecommand{\eprint}[2][]{\url{#2}}

\bibitem[{\citenamefont{Hasan and Kane}(2010)}]{Hasan10}
\bibinfo{author}{\bibfnamefont{M.~Z.} \bibnamefont{Hasan}} \bibnamefont{and}
  \bibinfo{author}{\bibfnamefont{C.~L.} \bibnamefont{Kane}},
  \bibinfo{journal}{Rev. Mod. Phys.} \textbf{\bibinfo{volume}{82}},
  \bibinfo{pages}{3045} (\bibinfo{year}{2010}).

\bibitem[{\citenamefont{Qi and Zhang}(2011)}]{Qi11}
\bibinfo{author}{\bibfnamefont{X.-L.} \bibnamefont{Qi}} \bibnamefont{and}
  \bibinfo{author}{\bibfnamefont{S.-C.} \bibnamefont{Zhang}},
  \bibinfo{journal}{Rev. Mod. Phys.} \textbf{\bibinfo{volume}{83}},
  \bibinfo{pages}{1057} (\bibinfo{year}{2011}).

\bibitem[{\citenamefont{Qi et~al.}(2008)\citenamefont{Qi, Hughes, and
  Zhang}}]{Qi08}
\bibinfo{author}{\bibfnamefont{X.-L.} \bibnamefont{Qi}},
  \bibinfo{author}{\bibfnamefont{T.~L.} \bibnamefont{Hughes}},
  \bibnamefont{and} \bibinfo{author}{\bibfnamefont{S.-C.} \bibnamefont{Zhang}},
  \bibinfo{journal}{Phys. Rev. B} \textbf{\bibinfo{volume}{78}},
  \bibinfo{pages}{195424} (\bibinfo{year}{2008}).

\bibitem[{\citenamefont{Fu and Kane}(2008)}]{Fu2008}
\bibinfo{author}{\bibfnamefont{L.}~\bibnamefont{Fu}} \bibnamefont{and}
  \bibinfo{author}{\bibfnamefont{C.~L.} \bibnamefont{Kane}},
  \bibinfo{journal}{Phys. Rev. Lett.} \textbf{\bibinfo{volume}{100}},
  \bibinfo{pages}{096407} (\bibinfo{year}{2008}).

\bibitem[{\citenamefont{Liu et~al.}(2009)\citenamefont{Liu, Liu, Xu, Qi, and
  Zhang}}]{Liu09}
\bibinfo{author}{\bibfnamefont{Q.}~\bibnamefont{Liu}},
  \bibinfo{author}{\bibfnamefont{C.-X.} \bibnamefont{Liu}},
  \bibinfo{author}{\bibfnamefont{C.}~\bibnamefont{Xu}},
  \bibinfo{author}{\bibfnamefont{X.-L.} \bibnamefont{Qi}}, \bibnamefont{and}
  \bibinfo{author}{\bibfnamefont{S.-C.} \bibnamefont{Zhang}},
  \bibinfo{journal}{Phys. Rev. Lett.} \textbf{\bibinfo{volume}{102}},
  \bibinfo{pages}{156603} (\bibinfo{year}{2009}).

\bibitem[{\citenamefont{Biswas and Balatsky}(2010)}]{Biswas10}
\bibinfo{author}{\bibfnamefont{R.~R.} \bibnamefont{Biswas}} \bibnamefont{and}
  \bibinfo{author}{\bibfnamefont{A.~V.} \bibnamefont{Balatsky}},
  \bibinfo{journal}{Phys. Rev. B} \textbf{\bibinfo{volume}{81}},
  \bibinfo{pages}{233405} (\bibinfo{year}{2010}).

\bibitem[{\citenamefont{Chen et~al.}(2010)}]{Chen10}
\bibinfo{author}{\bibfnamefont{Y.~L.} \bibnamefont{Chen}} \bibnamefont{et~al.},
  \bibinfo{journal}{Science} \textbf{\bibinfo{volume}{329}},
  \bibinfo{pages}{659} (\bibinfo{year}{2010}).

\bibitem[{\citenamefont{Schlenk et~al.}(2013)}]{Schlenk13}
\bibinfo{author}{\bibfnamefont{T.}~\bibnamefont{Schlenk}} \bibnamefont{et~al.},
  \bibinfo{journal}{Phys. Rev. Lett.} \textbf{\bibinfo{volume}{110}},
  \bibinfo{pages}{126804} (\bibinfo{year}{2013}).

\bibitem[{\citenamefont{She et~al.}(2013)\citenamefont{She, Fransson, Bishop,
  and Balatsky}}]{She13}
\bibinfo{author}{\bibfnamefont{J.-H.} \bibnamefont{She}},
  \bibinfo{author}{\bibfnamefont{J.}~\bibnamefont{Fransson}},
  \bibinfo{author}{\bibfnamefont{A.~R.} \bibnamefont{Bishop}},
  \bibnamefont{and} \bibinfo{author}{\bibfnamefont{A.~V.}
  \bibnamefont{Balatsky}}, \bibinfo{journal}{Phys. Rev. Lett.}
  \textbf{\bibinfo{volume}{110}}, \bibinfo{pages}{026802}
  (\bibinfo{year}{2013}).

\bibitem[{\citenamefont{Fransson et~al.}(2013)\citenamefont{Fransson, She,
  Pietronero, and Balatsky}}]{Fransson13}
\bibinfo{author}{\bibfnamefont{J.}~\bibnamefont{Fransson}},
  \bibinfo{author}{\bibfnamefont{J.~H.} \bibnamefont{She}},
  \bibinfo{author}{\bibfnamefont{L.}~\bibnamefont{Pietronero}},
  \bibnamefont{and} \bibinfo{author}{\bibfnamefont{A.~V.}
  \bibnamefont{Balatsky}}, \bibinfo{journal}{Phys. Rev. B}
  \textbf{\bibinfo{volume}{87}}, \bibinfo{pages}{245404}
  (\bibinfo{year}{2013}).

\bibitem[{\citenamefont{Anderson}(1975)}]{Anderson75}
\bibinfo{author}{\bibfnamefont{P.~W.} \bibnamefont{Anderson}},
  \bibinfo{journal}{Phys. Rev. Lett.} \textbf{\bibinfo{volume}{34}},
  \bibinfo{pages}{953} (\bibinfo{year}{1975}).

\bibitem[{\citenamefont{Varma}(1988)}]{Varma88}
\bibinfo{author}{\bibfnamefont{C.~M.} \bibnamefont{Varma}},
  \bibinfo{journal}{Phys. Rev. Lett.} \textbf{\bibinfo{volume}{61}},
  \bibinfo{pages}{2713} (\bibinfo{year}{1988}).

\bibitem[{\citenamefont{Micnas et~al.}(1990)\citenamefont{Micnas, Ranninger,
  and Robaszkiewicz}}]{Micnas90}
\bibinfo{author}{\bibfnamefont{R.}~\bibnamefont{Micnas}},
  \bibinfo{author}{\bibfnamefont{J.}~\bibnamefont{Ranninger}},
  \bibnamefont{and}
  \bibinfo{author}{\bibfnamefont{S.}~\bibnamefont{Robaszkiewicz}},
  \bibinfo{journal}{Rev. Mod. Phys.} \textbf{\bibinfo{volume}{62}},
  \bibinfo{pages}{113} (\bibinfo{year}{1990}).

\bibitem[{\citenamefont{Bar-Yam}(1991)}]{BarYam91}
\bibinfo{author}{\bibfnamefont{Y.}~\bibnamefont{Bar-Yam}},
  \bibinfo{journal}{Phys. Rev. B} \textbf{\bibinfo{volume}{43}},
  \bibinfo{pages}{359} (\bibinfo{year}{1991}).

\bibitem[{\citenamefont{Dzero and Schmalian}(2005)}]{Schmalian05}
\bibinfo{author}{\bibfnamefont{M.}~\bibnamefont{Dzero}} \bibnamefont{and}
  \bibinfo{author}{\bibfnamefont{J.}~\bibnamefont{Schmalian}},
  \bibinfo{journal}{Phys. Rev. Lett.} \textbf{\bibinfo{volume}{94}},
  \bibinfo{pages}{157003} (\bibinfo{year}{2005}).

\bibitem[{\citenamefont{Berg et~al.}(2008)\citenamefont{Berg, Orgad, and
  Kivelson}}]{Kivelson08}
\bibinfo{author}{\bibfnamefont{E.}~\bibnamefont{Berg}},
  \bibinfo{author}{\bibfnamefont{D.}~\bibnamefont{Orgad}}, \bibnamefont{and}
  \bibinfo{author}{\bibfnamefont{S.~A.} \bibnamefont{Kivelson}},
  \bibinfo{journal}{Phys. Rev. B} \textbf{\bibinfo{volume}{78}},
  \bibinfo{pages}{094509} (\bibinfo{year}{2008}).

\bibitem[{\citenamefont{Wachtel et~al.}(2012)\citenamefont{Wachtel, Bar-Yaacov,
  and Orgad}}]{Orgad12}
\bibinfo{author}{\bibfnamefont{G.}~\bibnamefont{Wachtel}},
  \bibinfo{author}{\bibfnamefont{A.}~\bibnamefont{Bar-Yaacov}},
  \bibnamefont{and} \bibinfo{author}{\bibfnamefont{D.}~\bibnamefont{Orgad}},
  \bibinfo{journal}{Phys. Rev. B} \textbf{\bibinfo{volume}{86}},
  \bibinfo{pages}{134531} (\bibinfo{year}{2012}).

\bibitem[{\citenamefont{Mal'shukov}(1991)}]{Malshukov91}
\bibinfo{author}{\bibfnamefont{A.~G.} \bibnamefont{Mal'shukov}},
  \bibinfo{journal}{Solid State Commun.} \textbf{\bibinfo{volume}{77}},
  \bibinfo{pages}{57} (\bibinfo{year}{1991}).

\bibitem[{\citenamefont{Gartstein and Mal'shukov}(1992)}]{Malshukov92}
\bibinfo{author}{\bibfnamefont{Y.~N.} \bibnamefont{Gartstein}}
  \bibnamefont{and} \bibinfo{author}{\bibfnamefont{A.~G.}
  \bibnamefont{Mal'shukov}}, \bibinfo{journal}{Solid State Commun.}
  \textbf{\bibinfo{volume}{83}}, \bibinfo{pages}{989} (\bibinfo{year}{1992}).

\bibitem[{\citenamefont{Vafek and Vishwanath}(2014)}]{Vafek14}
\bibinfo{author}{\bibfnamefont{O.}~\bibnamefont{Vafek}} \bibnamefont{and}
  \bibinfo{author}{\bibfnamefont{A.}~\bibnamefont{Vishwanath}},
  \bibinfo{journal}{Annu. Rev. Condens. Matter Phys.}
  \textbf{\bibinfo{volume}{5}}, \bibinfo{pages}{5.1} (\bibinfo{year}{2014}).

\bibitem[{\citenamefont{Wehling et~al.}(2014)\citenamefont{Wehling,
  Black-Schaffer, and Balatsky}}]{Wehling14}
\bibinfo{author}{\bibfnamefont{T.~O.} \bibnamefont{Wehling}},
  \bibinfo{author}{\bibfnamefont{A.~M.} \bibnamefont{Black-Schaffer}},
  \bibnamefont{and} \bibinfo{author}{\bibfnamefont{A.~V.}
  \bibnamefont{Balatsky}}, \bibinfo{journal}{Adv. Phys.}
  \textbf{\bibinfo{volume}{76}}, \bibinfo{pages}{1} (\bibinfo{year}{2014}).

\bibitem[{\citenamefont{Balatsky and Trugman}(1997)}]{Balatsky97}
\bibinfo{author}{\bibfnamefont{A.~V.} \bibnamefont{Balatsky}} \bibnamefont{and}
  \bibinfo{author}{\bibfnamefont{S.~A.} \bibnamefont{Trugman}},
  \bibinfo{journal}{Phys. Rev. Lett.} \textbf{\bibinfo{volume}{79}},
  \bibinfo{pages}{3767} (\bibinfo{year}{1997}).

\bibitem[{\citenamefont{Vojta}(2006)}]{Vojta0602}
\bibinfo{author}{\bibfnamefont{T.}~\bibnamefont{Vojta}}, \bibinfo{journal}{J.
  Phys. A: Math. Gen.} \textbf{\bibinfo{volume}{39}}, \bibinfo{pages}{R143}
  (\bibinfo{year}{2006}).

\bibitem[{\citenamefont{Spivak et~al.}(2008)\citenamefont{Spivak, Oreto, and
  Kivelson}}]{Spivak08}
\bibinfo{author}{\bibfnamefont{B.}~\bibnamefont{Spivak}},
  \bibinfo{author}{\bibfnamefont{P.}~\bibnamefont{Oreto}}, \bibnamefont{and}
  \bibinfo{author}{\bibfnamefont{S.~A.} \bibnamefont{Kivelson}},
  \bibinfo{journal}{Phys. Rev. B} \textbf{\bibinfo{volume}{77}},
  \bibinfo{pages}{214523} (\bibinfo{year}{2008}).

\bibitem[{\citenamefont{Nandkishore et~al.}(2013)\citenamefont{Nandkishore,
  Maciejko, Huse, and Sondhi}}]{Nandkishore13}
\bibinfo{author}{\bibfnamefont{R.}~\bibnamefont{Nandkishore}},
  \bibinfo{author}{\bibfnamefont{J.}~\bibnamefont{Maciejko}},
  \bibinfo{author}{\bibfnamefont{D.~A.} \bibnamefont{Huse}}, \bibnamefont{and}
  \bibinfo{author}{\bibfnamefont{S.~L.} \bibnamefont{Sondhi}},
  \bibinfo{journal}{Phys. Rev. B} \textbf{\bibinfo{volume}{87}},
  \bibinfo{pages}{174511} (\bibinfo{year}{2013}).

\bibitem[{\citenamefont{Schrieffer and Wolff}(1966)}]{Schrieffer66}
\bibinfo{author}{\bibfnamefont{J.~R.} \bibnamefont{Schrieffer}}
  \bibnamefont{and} \bibinfo{author}{\bibfnamefont{P.~A.} \bibnamefont{Wolff}},
  \bibinfo{journal}{Phys. Rev.} \textbf{\bibinfo{volume}{149}},
  \bibinfo{pages}{491} (\bibinfo{year}{1966}).

\bibitem[{\citenamefont{Taraphder and Coleman}(1991)}]{Coleman91}
\bibinfo{author}{\bibfnamefont{A.}~\bibnamefont{Taraphder}} \bibnamefont{and}
  \bibinfo{author}{\bibfnamefont{P.}~\bibnamefont{Coleman}},
  \bibinfo{journal}{Phys. Rev. Lett.} \textbf{\bibinfo{volume}{66}},
  \bibinfo{pages}{2814} (\bibinfo{year}{1991}).

\bibitem[{\citenamefont{Hewson}(1993)}]{Hewson}
\bibinfo{author}{\bibfnamefont{A.~C.} \bibnamefont{Hewson}},
  \emph{\bibinfo{title}{The Kondo Problem to Heavy Fermions}}
  (\bibinfo{publisher}{Cambridge University Press}, \bibinfo{year}{1993}).

\bibitem[{\citenamefont{Dobrosavljevi{\'c}
  et~al.}(1992)\citenamefont{Dobrosavljevi{\'c}, Kirkpatrick, and
  Kotliar}}]{Kotliar92}
\bibinfo{author}{\bibfnamefont{V.}~\bibnamefont{Dobrosavljevi{\'c}}},
  \bibinfo{author}{\bibfnamefont{T.~R.} \bibnamefont{Kirkpatrick}},
  \bibnamefont{and} \bibinfo{author}{\bibfnamefont{B.~G.}
  \bibnamefont{Kotliar}}, \bibinfo{journal}{Phys. Rev. Lett.}
  \textbf{\bibinfo{volume}{69}}, \bibinfo{pages}{1113} (\bibinfo{year}{1992}).

\bibitem[{\citenamefont{Zyuzin and Spivak}(1986)}]{Spivak86}
\bibinfo{author}{\bibfnamefont{A.~Y.} \bibnamefont{Zyuzin}} \bibnamefont{and}
  \bibinfo{author}{\bibfnamefont{B.~Z.} \bibnamefont{Spivak}},
  \bibinfo{journal}{JETP Lett.} \textbf{\bibinfo{volume}{43}},
  \bibinfo{pages}{234} (\bibinfo{year}{1986}).

\bibitem[{\citenamefont{Bulaevskii and Panyukov}(1986)}]{Bulaevskii86}
\bibinfo{author}{\bibfnamefont{L.~N.} \bibnamefont{Bulaevskii}}
  \bibnamefont{and} \bibinfo{author}{\bibfnamefont{S.~V.}
  \bibnamefont{Panyukov}}, \bibinfo{journal}{JETP Lett.}
  \textbf{\bibinfo{volume}{43}}, \bibinfo{pages}{240} (\bibinfo{year}{1986}).

\bibitem[{\citenamefont{Jagannathan et~al.}(1988)\citenamefont{Jagannathan,
  Abrahams, and Stephen}}]{Jagannathan88}
\bibinfo{author}{\bibfnamefont{A.}~\bibnamefont{Jagannathan}},
  \bibinfo{author}{\bibfnamefont{E.}~\bibnamefont{Abrahams}}, \bibnamefont{and}
  \bibinfo{author}{\bibfnamefont{M.~J.} \bibnamefont{Stephen}},
  \bibinfo{journal}{Phys. Rev. B} \textbf{\bibinfo{volume}{37}},
  \bibinfo{pages}{436} (\bibinfo{year}{1988}).

\bibitem[{\citenamefont{Shon and Ando}(1998)}]{Shon98}
\bibinfo{author}{\bibfnamefont{N.~H.} \bibnamefont{Shon}} \bibnamefont{and}
  \bibinfo{author}{\bibfnamefont{T.}~\bibnamefont{Ando}}, \bibinfo{journal}{J.
  Phys. Soc. Jpn.} \textbf{\bibinfo{volume}{67}}, \bibinfo{pages}{2421}
  (\bibinfo{year}{1998}).

\bibitem[{\citenamefont{Tkachov and Hankiewicz}(2011)}]{Tkachov11}
\bibinfo{author}{\bibfnamefont{G.}~\bibnamefont{Tkachov}} \bibnamefont{and}
  \bibinfo{author}{\bibfnamefont{E.~M.} \bibnamefont{Hankiewicz}},
  \bibinfo{journal}{Phys. Rev. B} \textbf{\bibinfo{volume}{84}},
  \bibinfo{pages}{035444} (\bibinfo{year}{2011}).

\bibitem[{\citenamefont{Balatsky et~al.}(2006)\citenamefont{Balatsky, Vekhter,
  and Zhu}}]{BalatskyRMP}
\bibinfo{author}{\bibfnamefont{A.~V.} \bibnamefont{Balatsky}},
  \bibinfo{author}{\bibfnamefont{I.}~\bibnamefont{Vekhter}}, \bibnamefont{and}
  \bibinfo{author}{\bibfnamefont{J.-X.} \bibnamefont{Zhu}},
  \bibinfo{journal}{Rev. Mod. Phys.} \textbf{\bibinfo{volume}{78}},
  \bibinfo{pages}{373} (\bibinfo{year}{2006}).

\bibitem[{\citenamefont{Ostrovsky et~al.}(2006)\citenamefont{Ostrovsky, Gornyi,
  and Mirlin}}]{Mirlin06}
\bibinfo{author}{\bibfnamefont{P.~M.} \bibnamefont{Ostrovsky}},
  \bibinfo{author}{\bibfnamefont{I.~V.} \bibnamefont{Gornyi}},
  \bibnamefont{and} \bibinfo{author}{\bibfnamefont{A.~D.}
  \bibnamefont{Mirlin}}, \bibinfo{journal}{Phys. Rev. B}
  \textbf{\bibinfo{volume}{74}}, \bibinfo{pages}{235443}
  (\bibinfo{year}{2006}).

\bibitem[{\citenamefont{Gonzalez and Perfetto}(2008)}]{Gonzalez08}
\bibinfo{author}{\bibfnamefont{J.}~\bibnamefont{Gonzalez}} \bibnamefont{and}
  \bibinfo{author}{\bibfnamefont{E.}~\bibnamefont{Perfetto}},
  \bibinfo{journal}{J. Phys.: Condens. Matter} \textbf{\bibinfo{volume}{20}},
  \bibinfo{pages}{145218} (\bibinfo{year}{2008}).

\bibitem[{\citenamefont{Sigrist and Rice}(1995)}]{Sigrist95}
\bibinfo{author}{\bibfnamefont{M.}~\bibnamefont{Sigrist}} \bibnamefont{and}
  \bibinfo{author}{\bibfnamefont{T.~M.} \bibnamefont{Rice}},
  \bibinfo{journal}{Rev. Mod. Phys.} \textbf{\bibinfo{volume}{67}},
  \bibinfo{pages}{503} (\bibinfo{year}{1995}).

\bibitem[{\citenamefont{Keles et~al.}()\citenamefont{Keles, Andreev, Kivelson,
  and Spivak}}]{Spivak14}
\bibinfo{author}{\bibfnamefont{A.}~\bibnamefont{Keles}},
  \bibinfo{author}{\bibfnamefont{A.~V.} \bibnamefont{Andreev}},
  \bibinfo{author}{\bibfnamefont{S.~A.} \bibnamefont{Kivelson}},
  \bibnamefont{and} \bibinfo{author}{\bibfnamefont{B.~Z.}
  \bibnamefont{Spivak}}, \bibinfo{note}{arXiv:1405.7090 [cond-mat.supr-con]}.

\bibitem[{\citenamefont{Teitel and Jayaprakash}(1983)}]{Teitel8302}
\bibinfo{author}{\bibfnamefont{S.}~\bibnamefont{Teitel}} \bibnamefont{and}
  \bibinfo{author}{\bibfnamefont{C.}~\bibnamefont{Jayaprakash}},
  \bibinfo{journal}{Phys. Rev. B} \textbf{\bibinfo{volume}{27}},
  \bibinfo{pages}{598} (\bibinfo{year}{1983}).

\bibitem[{\citenamefont{Korshunov}(2006)}]{Korshunov06}
\bibinfo{author}{\bibfnamefont{S.~E.} \bibnamefont{Korshunov}},
  \bibinfo{journal}{Phys.-Usp.} \textbf{\bibinfo{volume}{49}},
  \bibinfo{pages}{225} (\bibinfo{year}{2006}).

\bibitem[{\citenamefont{Hasenbusch et~al.}(2005)\citenamefont{Hasenbusch,
  Pelissetto, and Vicari}}]{Vicari05}
\bibinfo{author}{\bibfnamefont{M.}~\bibnamefont{Hasenbusch}},
  \bibinfo{author}{\bibfnamefont{A.}~\bibnamefont{Pelissetto}},
  \bibnamefont{and} \bibinfo{author}{\bibfnamefont{E.}~\bibnamefont{Vicari}},
  \bibinfo{journal}{J. Stat. Mech.} \textbf{\bibinfo{volume}{2005}},
  \bibinfo{pages}{P12002} (\bibinfo{year}{2005}).

\bibitem[{\citenamefont{Alba et~al.}(2010)\citenamefont{Alba, Pelissetto, and
  Vicari}}]{Vicari10}
\bibinfo{author}{\bibfnamefont{V.}~\bibnamefont{Alba}},
  \bibinfo{author}{\bibfnamefont{A.}~\bibnamefont{Pelissetto}},
  \bibnamefont{and} \bibinfo{author}{\bibfnamefont{E.}~\bibnamefont{Vicari}},
  \bibinfo{journal}{J. Stat. Mech.} \textbf{\bibinfo{volume}{2010}},
  \bibinfo{pages}{P03006} (\bibinfo{year}{2010}).

\bibitem[{\citenamefont{Metlitski et~al.}()\citenamefont{Metlitski, Kane, and
  Fisher}}]{Metlitski13}
\bibinfo{author}{\bibfnamefont{M.~A.} \bibnamefont{Metlitski}},
  \bibinfo{author}{\bibfnamefont{C.~L.} \bibnamefont{Kane}}, \bibnamefont{and}
  \bibinfo{author}{\bibfnamefont{M.~P.~A.} \bibnamefont{Fisher}},
  \bibinfo{note}{arXiv:1306.3286 [cond-mat.str-el]}.

\bibitem[{\citenamefont{Bonderson et~al.}(2013)\citenamefont{Bonderson, Nayak,
  and Qi}}]{Bonderson13}
\bibinfo{author}{\bibfnamefont{P.}~\bibnamefont{Bonderson}},
  \bibinfo{author}{\bibfnamefont{C.}~\bibnamefont{Nayak}}, \bibnamefont{and}
  \bibinfo{author}{\bibfnamefont{X.-L.} \bibnamefont{Qi}}, \bibinfo{journal}{J.
  Stat. Mech.} p. \bibinfo{pages}{P09016} (\bibinfo{year}{2013}).

\bibitem[{\citenamefont{Wang et~al.}(2013)\citenamefont{Wang, Potter, and
  Senthil}}]{Senthil13}
\bibinfo{author}{\bibfnamefont{C.}~\bibnamefont{Wang}},
  \bibinfo{author}{\bibfnamefont{A.~C.} \bibnamefont{Potter}},
  \bibnamefont{and} \bibinfo{author}{\bibfnamefont{T.}~\bibnamefont{Senthil}},
  \bibinfo{journal}{Phys. Rev. B} \textbf{\bibinfo{volume}{88}},
  \bibinfo{pages}{115137} (\bibinfo{year}{2013}).

\bibitem[{\citenamefont{Chen et~al.}(2014)\citenamefont{Chen, Fidkowski, and
  Vishwanath}}]{Vishwanath13}
\bibinfo{author}{\bibfnamefont{X.}~\bibnamefont{Chen}},
  \bibinfo{author}{\bibfnamefont{L.}~\bibnamefont{Fidkowski}},
  \bibnamefont{and}
  \bibinfo{author}{\bibfnamefont{A.}~\bibnamefont{Vishwanath}},
  \bibinfo{journal}{Phys. Rev. B} \textbf{\bibinfo{volume}{89}},
  \bibinfo{pages}{165132} (\bibinfo{year}{2014}).

\end{thebibliography}

\end{document}